# Joint Activity Design Heuristics for Enhancing Human-Machine Collaboration

Mohammadreza Jalaeian, Dane A. Morey, and Michael F. Rayo

*Abstract*—Joint activity describes when more than one agent (human or machine) contributes to the completion of a task or activity. Designing for joint activity focuses on explicitly supporting the interdependencies between agents necessary for effective coordination among agents engaged in the joint activity. This builds and expands upon designing for usability to further address how technologies can be designed to act as effective team players. Effective joint activity requires supporting, at minimum, five primary macrocognitive functions within teams: Event Detection, Sensemaking, Adaptability, Perspective-Shifting, and Coordination. Supporting these functions is equally as important as making technologies usable. We synthesized fourteen heuristics from relevant literature including display design, human factors, cognitive systems engineering, cognitive psychology, and computer science to aid the design, development, and evaluation of technologies that support joint human-machine activity.

## I. Introduction

Human-Machine interactions have evolved and matured from the earliest days when machines were used for simple work. Recent advances in Artificial Intelligence (AI) and Machine Learning (ML) technologies have accelerated human-machine interactions progressing from simple tool-based engagements to complex cognitive collaborations [1]. Machines are being designed to perform an increasing set of functions and are being expected to engage more deeply in the collaborative joint activities related to these functions. This shift in machine capabilities and expectations demands a corresponding re-evaluation and broadening of design and evaluation principles to support joint human-machine activity in ways that lie outside the boundaries of traditional usability methods and models [2].

Traditional usability heuristics, such as those proposed by [3], provide a strong foundation focusing primarily on surface-level interactions such as enhancing the ease of use, efficiency, and satisfaction in human-machine interaction. These heuristics are primarily oriented towards actions and responses but offer limited support for the essential macrocognitive functions associated with effective teamwork including event detection, sensemaking, adaptability, perspective shifting, and coordination, all of which are vital in the close collaboration of humans and machines with joint activities [2], [4], [5], [6]. These heuristics are primarily oriented towards actions and responses but offer limited support for the essential macrocognitive functions associated with effective teamwork including event detection, sensemaking, adaptability, perspective shifting, and coordination. All of these macrocognitive functions are vital in the close collaboration of humans and machines with joint activities in high-stakes and dynamic environments with little room for error [2], [5]. This reliance on macrocognitive functions is evident in domains where the ability to process complex information and adapt to changing conditions is crucial. For example, in commercial aviation, pilots must not only manage automated systems but also integrate diverse streams of information, recognize potential issues early, and make swift, informed decisions to ensure passenger safety [7], [8]. Similarly, in healthcare, where errors can have severe consequences on patient outcomes, macrocognitive functions enable clinicians to prioritize tasks, synthesize patient data, and adapt treatments based on emerging patterns [9].

Although theoretical research [4], [10], [11], [12, p. 201], [13], [14], [15] provides robust frameworks for understanding human-machine interaction in joint human-machine cognitive systems, there remains a significant gap in applying these frameworks practically. Theoretical insights often remain restricted to academic literature and beyond the reach of practitioners who have the practical task of designing and implementing these technologies in operational settings. With the lack of appropriate methods and tools that can be used to design and evaluate such joint human-machine cognitive systems, practitioners who are responsible for designing and deploying these technologies in natural settings may face challenges [16], [17].

Recognizing these limitations and building upon our prior work [18], we propose a set of fourteen heuristics as a response to these challenges, narrowing the gap between science and practice to meet the demands of this evolving technology landscape. These heuristics, which serve as an extension to the traditional usability heuristics, have been developed out of a synthesis between cognitive systems engineering principles, usability, human-computer interactions, and practical design needs to produce technologies that support both usability and the macrocognitive functions that enable joint cognitive work. Our proposed fourteen heuristics integrate macrocognitive functions into joint human-machine cognitive systems, aiming to enhance efficiency and performance while positioning technology as a supportive tool rather than a replacement. In addition to contextualizing these theoretical guidelines, the heuristics presented were revised based on feedback from practical application to better operationalize them for use. The aim of these heuristics is to embed strong scientific findings into practical tools, providing an impactful toolset for the community of designers and engineers tasked with creating more effective, resilient, and intuitive human-machine systems [19].

These new heuristics are an important step forward for the design and early evaluation of joint human-machine cognitive systems. They are created to reflect the challenges of human-machine collaboration to ensure that machines work not only as tools but as team players [20]. This pragmatic synthesis aims to both extend the theoretical work in cognitive systems engineering and also operationalize these insights for designers and operators to create more effective and harmonious human-machine interactions.

## II. METHODS

We expanded our prior research [18] to synthesize a set of pragmatic heuristics for joint activity design. The process started with an extensive review of the literature including 120 articles to identify critical challenges, existing gaps, and current guidance in the design and implementation of joint human-machine cognitive systems. Topic areas included display design, human factors, cognitive systems engineering, cognitive psychology, and computer science. This literature review was used to synthesize a set of machine-focused requirements for supporting joint activity. These requirements are published elsewhere [18].

These requirements were presented to a set of practitioners to evaluate their practicality and effectiveness as a preliminary set of heuristics. We collected feedback from eight practitioners utilizing these heuristics in realistic settings to evaluate a total of five designs. Practitioners had a wide range of human factors or design experience ranging from one to fifteen years. From this feedback, the set of preliminary heuristics were revised.

The updated set of heuristics was then given to a new set of thirteen practitioners with a wide range of one to 25 years of human factors or design experience. Again, their feedback on the heuristics and the process of using them to evaluate technologies was collected. They were trained on the intended heuristic evaluation process. Each participant used the heuristics to evaluate two technologies. Participants could interact with the technologies as they would in a normal usage setting. Participants rated the application of the heuristics to each technology on a five-point Likert scale from 1 (clearly violated) to 5 (clearly applied). For each heuristic, participants provided a brief written rationale for their scoring. They were also asked to give written comments on the language of each heuristic to obtain qualitative feedback. Practitioners' heuristic scoring, reasoning behind that scoring, and post-evaluation feedback were analyzed through a thematic qualitative approach. The analysis provided common themes and patterns in responses recommending further revisions. All feedback from such evaluations was collected and carefully analyzed for another round of revisions on all heuristics.

The following sections present a new set of fourteen heuristics resulting from our initial synthesis of the literature and two rounds of practitioner feedback. These heuristics are organized around five core macrocognitive functions [4]: event detection (D), sensemaking (S), adaptability (A), perspective-shifting (P), and coordination (C) . Heuristics related to each macrocognitive function are described in separate sections. The title of heuristic is labeled with a letter (D, S, A, P, or C) and number (1-14) to indicate the macrocognitive grouping and index within the set, respectively. Following the title, a one-sentence summary of each heuristic is included in bolded text.

## III. HEURISTICS TO SUPPORT JOINT EVENT DETECTION

Joint event detection is a prerequisite for well-informed decision-making in dynamic environments. It refers to the identification and interpretation of important changes or occurrences within an environment that might affect subsequent actions and decisions. Effective event detection demands an ability to discriminate between information and background noise, which forms an integral process for maintaining high levels of situation awareness [21]. Event detection is a crucial feature of cognitive systems engineering to be capable of developing systems that enhance human cognitive processes to improve recognition and response accuracy [22], [23]. Without it, the subsequent functions of sensemaking and responding/taking action cannot and will not occur [24]. Therefore, it is important to design displays that could support effective event detection to enable improvement of joint cognitive systems that can handle the demands of modern operational environments. This section considers the development of heuristics that support system improvement to allow joint event detection including identifiable and distinguishable signals and emergent patterns. We examine the principles and applications that underpin these heuristics, illustrating how they can be implemented to improve clarity and reduce cognitive load for users. The heuristics suggest strategies to optimize how signals are presented to maximize their complementarity with human perceptual systems, thus improving the overall performance of joint cognitive systems.

### A. Identifiable and Distinguishable Signals (D1)

**Displays should present all relevant signals to be identifiable and distinguishable by users, even in noisy and overcrowded environments.** Signals (e.g., data on a display, notifications and alerts, important trends in graphs, alarms) do not exist in isolation; therefore, they must be designed in the context of and mitigate the likelihood of masking from the other signals that can be present [25], [26], [27]. Users and displays can become saturated by an overwhelming volume of signals, many of which are often irrelevant or redundan [28], [29]. In these noisy environments, just because a signal is present does not mean that it will be identified, perceived, or understood. This signal overload has the potential to distract, disguise, overwhelm, or otherwise obscure users from distinguishing relevant from irrelevant signals [30]. Signals are considered "identifiable" when they clearly stand out from the background and can be quickly recognized by users, such as a red flashing light and accompanying auditory alarm on a car's dashboard being correctly identified as check engine alarm [31]. "Distinguishable" signals are those that are differentiable from other signals in that environment, like recognizing the difference between email notification sounds and calendar reminders.

To reduce signal overload and masking and to ensure important signals are discernible from irrelevant ones, we recommend the following strategies:

1) *Complex and Heterogeneous Signals*: Employ a blend of characteristics such as color, shape, pattern, timbre, and roughness to make critical signals stand out in multiple dimensions [27]. For example, dynamic animations could be used for critical notifications, in contrast to static icons for less urgent information.
2) *Different Sensory Modalities*: Utilize different senses (e.g., visual, auditory) to prevent signals from competing with one another and to reduce the potential for one signal to mask another.
3) *Unique Perceptual Characteristics*: Design each signal with distinct qualities not shared with others, like a unique sound for a high-priority message, to enhance identifiability.
4) *Grouping Similar Signals*: Cluster similar signals using some shared perceptual characteristics (e.g., color or timbre) while varying other perceptual characteristics (e.g., pattern or intensity) to maintain uniqueness [27].

*B. Emergent Patterns (D2)*

**Data should be organized to form emergent patterns that equip users to quickly recognize events.** Effective displays produce emergent perceptual patterns for both expected and unexpected events [32], [33]. Emergent patterns are identifiable data signatures that occur when individual display components are perceived together to emphasize their overall significance or meaning. An emergent pattern in the financial domain could be, for instance, a gradual increase in the trading volume of a stock, indicating a consistent growth of the company in the market. This signature of consistent company growth is not obvious from the stock value of any individual day, which is likely to be noisy day-to-day, but rather a signature that emerges from seeing the general trend over a longer period of time. If the organization of data closely matches relationships in the world, the patterns which emerge for unexpected events should appear perceptually different than the patterns which emerge for expected events [27], [34]. Users should be able to quickly recognize anomalies from the perceptual patterns emerging from the display, even if it is uncertain what the anomaly is. Over time, users should be able to quickly recognize recurring events based on the perceptual patterns present in the display. This is how users build expertise. To facilitate this process, displays must be structured so that:

1) *Important data relationships become perceptually prominent*: This is achieved by organizing lower-order data (i.e., the details) in a way that higher-order phenomena (i.e., the meaning) are intuitively evident and stand out in the display. This approach allows users to maintain the correct frame of reference and enhances the recognition of expected events based on their familiarity with these relationships [10], [35].
2) *Users can keep pace with the tempo of work*: The display should allow users to detect these patterns at a pace that matches the requirements and dynamics of their work [36]. If users must respond quickly, they must be able to accurately detect patterns even more quickly. Often, emphasizing changes from the established baseline is an effective strategy to flag potential issues [36]. Recognizing changes from baseline is crucial for identifying both expected and unexpected events. By emphasizing this change, users can detect anomalies and novel events that may signify important developments or require immediate attention.

Designing displays in a way that keeps essential data relationships perceptually salient and intuitive means that users can keep an appropriate frame of reference and adapt to the pace required for their particular operational context in order to maximize the utility of the emergent patterns in these displays. By making salient significant data relationships and changes from baseline, displays can help users recognize and react to both expected and unexpected events in a timely manner.

IV. HEURISTICS TO SUPPORT JOINT SENSEMAKING

Sensemaking is the process by which users come to an understandable description of the situation that enables appropriate action in the light of emerging circumstances and is another critically important macrocognitive function [37]. It ensures that both human operators and automated systems can interpret data in a coherent and meaningful manner. The integration of sensemaking to the design of displays thus supports human-machine interactions that are more attuned to human cognitive strategies by providing clear and context-rich information [20], [38]. Such systems support joint cognitive activities by aligning human reasoning patterns and machine processes to foster a shared understanding for enhancing performance and collaboration.

One of the design issues for sensemaking in joint activities is how to display data that can be readily interpreted by people even in situations which were unanticipated by designers. This calls for display design that will not just display data but frame it into understandable contexts the same way humans use cognitive processing [37], which supports better understanding of complex situations where fast comprehension and decision-making are necessary. In the following sections, we propose three heuristics that support integration of sensemaking into the design of joint cognitive systems.

*A. Display Dimensions Matching Real-World Complexity and Uncertainty (S3)*

**Displays should present all the important data relationships needed to understand the world in a way that matches real-world complexities and dynamics.** Displays are often a user's window to the world. This window needs to accurately depict all important aspects and data relationships in the world for a user to understand what is happening [36]. Displays should reflect the analyses that have determined what these aspects and relationships are [39]. Additionally, the way data relationships in the world are depicted in the display should be influenced by the degree of uncertainty in the world [40].

In low-uncertainty, low-complexity situations, data representation strategies that are more precise and compact are often advantageous. In high-uncertainty, high complexity situations, data representations strategies that provide more context and data relationships are often advantageous (see TABLE 1). For dynamic worlds, a mix of display strategies or a dynamic display is likely merited [36], [41], [42].

TABLE I
DATA DISPLAY STRATEGIES BASED ON ENVIRONMENTAL COMPLEXITY AND UNCERTAINTY

| Low-Uncertainty, Low-Complexity | High-Uncertainty, High-Complexity |
|---|---|
| Numbers, icons, earcons | Graphs |
| Categorical data | Continuous data |
| Displays which are triggered or pop-up by some condition | Displays which are always present |

Environments characterized by low complexity and uncertainty generally favor data display methods such as straightforward numerical indicators, icons, and earcons (i.e., the auditory equivalent of a visual icon) [39], [43]. These simple categorical data representations usually appear on the displays when certain conditions are met. However, in high-uncertainty, high-complexity environments, a more dynamic and continuous approach to data representation is advantageous. Within such settings, graphs, sonifications [44] and other analogical, continuous data representations can offer a more complete overview of changes within variables. Under such conditions, important information is always available on the display. These differences in the display strategy are essential for designing the joint cognitive system to effectively support the needs of different operational contexts, with respect to both usability and decision-making performance.

*B. Data in Context (S4)*

**Displays should set present information in the context of past information, future expectations, and other related information.** Data is rarely meaningful in isolation. Putting data in context refers to framing current information within the backdrop of related data points to reveal a more comprehensive narrative [36]. Context includes past values or future predictions of a data variable, a data point's relationship to a different variable, or any other relevant information that can influence the interpretation of the current data. Often, putting data in context can convey to people more information about a signal than the signal was explicitly designed to convey, like the relevance of the signal to the current situation [45]. For example, marathon runners can have resting heart rates that are below alarm thresholds for many hospitals, causing alarms to constantly go off. Setting these alarms in the context of a marathon runner's resting heart rate can reveal to people that the low alarm thresholds are irrelevant or miscalibrated for such a patient.

Context provides the background into which individual data points fit, offering insights into patterns, trends, and potential outcomes [36], [46], [47]. Setting current data in the context of past data provides additional information about the direction and speed at which the system is changing. Setting current data in the context of future expectations provides additional information about potential anomalies and how well projections of future state match reality [48]. Setting current data in the context of other related data provides additional information about how data might relate to one another and larger systemic patterns. Some strategies for visualizing data in context include:

1) *Annotations and Baselines*: Use annotations to establish contextual baselines that serve as reference points for understanding current data.
2) *Dynamic Highlighting*: Emphasize changes over time (direction and speed) and rate of change (acceleration) by highlighting the past data informing the extrapolation and changes in predictions over time.
3) *Visual Cues for Anomalies*: Implement design strategies that make the data more informative by perceptually signaling when data deviates from expected patterns, thus identifying the presence of an anomaly.

For example, in a public health tracking dashboard used by a city's health department, the current number of flu cases is presented not just as a static number but contextualized with the past data to compare this year's flu cases to the same period in previous years to detect patterns or anomalies in the occurrence of flu cases. Current trends are used to forecast future flu cases and prepare the healthcare system accordingly, such as by increasing vaccine supplies or opening more flu clinics. Evaluating the current flu cases in conjunction with other health indicators, like hospital admissions or school absenteeism, can be used to assess the broader impact of the flu season on the community. By visually integrating this context, the health department can make informed decisions, anticipate needs, and efficiently allocate resources to manage the flu season effectively.

V. HEURISTICS TO SUPPORT HUMAN ADAPTABILITY

In a joint human-machine system, human adaptability is responsible for flexibility and resilience within dynamically changing, unpredictable conditions [15], [42]. A person's ability to change behavior and strategies to respond to a changing environment or in anticipation of potential future states is part of the function [22]. To support human adaptability, a display design should account for adjusting to a user's needs in real-time to allow effective collaboration between human and machine [46]. To achieve this goal, we introduce three heuristics: utilize the minimum viable technology assertiveness, defer to people, and modulate involvement in automated systems. These heuristics help designers create adaptable and user-centric systems that account for and enhance human decision-making authority, allowing safety and efficiency to remain high while reducing the risk of adverse outcomes due to overly assertive or unadaptable automated responses.

*A. Minimum Viable Technology Assertiveness (A5)*

**Displays should be conservative with the assertiveness of technology interjections, weighing the dangers of**

**misleading disruptions.** This heuristic emphasizes that a display should minimize technological assertiveness to prevent unnecessary interruptions and minimize the costs of misleading interruptions [49], [50], [51]. How strongly a display asserts or interjects notifications, recommendations, alerts, or other information to people can influence and/or disrupt people's decision-making and the actions they take. Recommendations have been demonstrated to change people's problem-solving processes, for better and for worse [52]. In high-stakes environments, a higher level of disruptiveness may seem appropriate to ensure critical information captures attention. However, this must be carefully managed to avoid misleading users or inducing suboptimal decisions. When technology interjections are highly assertive but misleading, they can substantially degrade human performance [24], [36], [52]. Table 2 below gives examples of different degrees of technology assertiveness, which are augmented from [2]. Please note that technologies that filter information or otherwise make an inference, which include artificial intelligence and all other computational analytics solutions, are designated as being at least moderately assertive, which brings with it the associated risks mentioned above.

TABLE II
DEGREES OF ASSERTIVENESS IN A DISPLAY

| Assertiveness | Examples |
|---|---|
| Most Assertive | **Takeover:** The technology takes control from the user and directly takes actions without input from the user. |
| | **Hard Stop:** The technology completely stops a user's workflow and does not permit work to continue until some criteria are met. |
| | **Interrupting:** The technology interrupts a user's workflow and requires some acknowledgment before work is permitted to continue. |
| Moderately Assertive | **Recommending:** The technology offers a suggestion, filters, or removes what it thinks is irrelevant, or otherwise asserts an inference. |
| Least Assertive | **Highlighting:** The technology highlights or draws the user's attention to what information it thinks is important. |
| | **Displaying:** The technology displays information but remains neutral in what information is important or irrelevant. |
| | **Nothing:** The technology does nothing and displays nothing. |

*B. Defer to People (A6)*

**Users are responsible for system outputs, so users must be able to take full control of the automated system.** To defer to people means that users should be granted full control and automation must defer to and follow peoples' directions [20], [46]. This is true not only from the perspective of people being able to override their automated counterparts, but also that the automation's outputs do not overly influence or override peoples' judgments. Good teammates can direct and be directed by each other [53]. Automation and displays can (and will) influence the way people think. However, people are always responsible for system outcomes (even when the automation is in control) [54], so users must be able to take full control of the automated system from both perspectives.

Deferring to people implies that automated systems must be designed to acknowledge and facilitate human decision-making authority. The systems should support and enhance human control rather than supplant it. This deference is important, especially when system outputs affect critical outcomes or when automated decisions may need to be overridden by users' judgment. Automation which does not defer to people can have catastrophic consequences. For example, the automation suite on board the Boeing 737 Max continued to push the nose of the aircraft down, despite persistent efforts of the pilots to retake control and override the automation [55]. As a result, the automation crashed two airplanes and killed a total of 346 people.

Two categories of user oversight include human override and mitigating undue automation influence. The former requires that systems must always allow for user override. This ensures that in any situation, regardless of the system's automated decision-making processes, users retain ultimate control and can intervene as necessary. The latter requires building aids to help users break away from the inevitable influence of automation when it is incorrect or misleading [56].

Deferring to human input encompasses, but is not limited to, manual override capabilities. While manual override is a mechanism that allows users to directly control the system, deferring to people is broader. It includes designing systems that are receptive to human input at various levels, ensuring that automated responses do not override human commands and that there is a provision for dialogue between the user and the system. It's about creating a mutual relationship where the overall system is an extension of the human operator.

*C. Modulating Involvement in Automated Systems (A7)*

**The display and workload should support users' ability, at the required tempo, to increase and decrease how much they actively participate or contribute to the actions of the automated system.** The display and workload constraints should not force people to rely on automation to achieve their objectives [57]. This implies the task should be something that people could accomplish, if necessary, without fully utilizing the automation. If people are so overloaded with other tasks that they cannot contribute to the automation's actions, they will be held responsible for automated actions they cannot possibly control given their current demands. People must have both the ability and the time to monitor the automation and intervene when necessary [24]. Varying levels of involvement refer to the degree to which users interact with, direct, and oversee automated systems. This can range from full control and direct action to passive monitoring or even temporary disengagement when appropriate. A well-designed joint cognitive system provides cues to people to support effective and fluid modulation of their involvement [58]. The system should facilitate this spectrum of involvement, with deliberately designed features that produce signals (implicit or explicit) that convey to users when to take charge or step back [59].

VI. HEURISTICS TO SUPPORT JOINT PERSPECTIVE-SHIFTING

Joint perspective-shifting is a critical macrocognitive function that allows users to effectively navigate and integrate complex information in joint human-machine systems. Designing displays for effective joint perspective shifting is important in environments where decision making depends on the integration of diverse, often disparate, sets of information. It requires displays to not only present organized data but also allows for moving between different levels of data analysis without losing the context or creating too much cognitive load [60], [61]. Displays that are designed to support joint perspective shifting allow people to switch between different sets of information, understand the relationships among them, and synthesize them into a coherent whole. This process enables a person to detect anomalies and make decisions with a more holistic understanding of the dynamic situation making them more effective in managing complex environments. The following two heuristics help design displays that support joint perspective shifting in human-machine systems.

*A. Big Picture and Details (P8)*

**Displays should organize—but not reduce—data to reveal the big picture and details simultaneously.** Well-organized data enables users to seamlessly shift perspectives [62]. This strategy should be employed for every concept that the technology presents. A dual view should be maintained without diminishing the richness of data, allowing users to quickly determine the gist of each display without necessarily having to know the details [36].

The *big picture* refers to the overarching patterns that emerge from synthesizing data, while *details* pertain to the specific, often granular information that makes up these larger patterns. Displays should support the recursive ability to process more concrete data or primitives (i.e., signals) to be able to perceive (i.e., detect and interpret) larger groupings that represent higher order phenomena. Those groupings then become primitives to perceive higher-order phenomena (i.e., bigger picture).

*B. Fluidly Shifting Across Multiple Perspectives (P9)*

**Interactions with the automated system should enable and encourage fluid extraction and integration of information across multiple perspectives and/or spaces.** The design of automated systems should foster the fluid navigation and integration of information from various perspectives and contextual spaces. Recognizing that no single viewpoint can include all critical insights in complex systems, the display should enable users to effortlessly shift, compare, and synthesize diverse perspectives to form a comprehensive understanding of the situation. Shifting perspectives can be disorienting [60]. It can be difficult for people to reorient to the new perspective, extract information from the new perspective, remember the prior perspective, and compare across multiple perspectives. Displays should make shifting perspectives as fluid as possible. Each perspective people can access should provide data about the location of the perspective in relation to others to help people build a mental map reflective of the automated process and real work environment. The transition between perspectives should depict how the two perspectives relate to each other to minimize the costs of reorienting to a new perspective.

To support this, the display configuration should establish strong cues that support users' easy transitions between perspectives (i.e., frames of reference). It should give cues about the relational context of each perspective and help users build a mental model that reflects the structure of the automated system and the real-world scenario it represents [60]. Transitions between perspectives should be designed in a way that shows the interrelation of information to reduce the cognitive load from taking on a new perspective. The display naturally shows how one perspective relates to another; as a result, users experience less effort in transitioning between different levels of data analysis, which increases user adaptability and comprehension of information presented in the display (i.e., visual momentum, [56]).

VII. HEURISTICS TO SUPPORT HUMAN-MACHINE COORDINATION

Human-machine coordination is a significant macrocognitive function, as it facilitates an effective interaction between humans and the automated system to ensure both are working towards a common goal. The function is supported by heuristics which enhance transparency and explainability of automated systems for adequate understanding and anticipation of machine behaviors by users [26]. Furthermore, displays should be designed with a view to communicate the transitions more clearly between different modes and operational states of the automation, as these fundamentally influence the expectations of the user and how they interact with the automation [63]. When these systems are well coordinated, they become more likely to be accepted and trusted by users, which results in the desired effective and efficient outcomes in complex operational environments [5]. However, effective coordination must also reveal situations of misalignment and opportunities for realignment [64]. In the following sections, we discuss five heuristics that support coordination as a macrocognitive function in human-machine systems.

*A. Global Explanations for Machine Decision Making (C10)*

**Displays should include global explanations to allow users to understand what data is accessible to the automation, how inputs are calculated from this data, and how different data and input features are transformed into outputs.** Global explanations clarify how input data is generally transformed to an output to reveal the underlying logic and possible biases by such systems. Such transparency is necessary for users to be able to evaluate the system's reliability and to understand conditions under which it might fail [26]. In alignment with AI principles of global explanation, displays should clearly present how the automated system generally transforms inputs into outputs. This includes what data is accessible to the automated system, how the automated system weighs different input features or instances, and the general rules by which it calculates outputs from these inputs. This also includes circumstances or combinations of inputs which cause the automated system to fail or produce erroneous outputs [65]. This information helps users understand the conditions under which the automation is likely to perform well or poorly.

The display design should enable users to understand more about the processes that the automated system goes through while making decisions through explanations, transparent data, and the ability to explore and even interact with the underlying reasoning of the system. Different data are sometimes available to humans and automated systems. The display should show what data makes up the view of the automated system including data available to both people and the automation, data available only to people, data available only to the automation, and data available neither to people nor automation. This information should help users develop a sense of the system's limitation and of what the system does know.

Even when both humans and automated systems are provided with the same data, automated systems often do not process the data in the same way as humans. The display should explain how an automated system makes computations of the inputs to its model from available data, letting users understand the machine's internal logic. In addition, displays that allow the users to have a better understanding of how the automated system works typically let a person explore how the automated system's outputs would be changed if the input had been changed. This includes manipulation of input features to construct failure scenarios.

*B. Local Explanations for Machine Decision Making (C11)*

**Displays should reveal how and why an automated system is making decisions for the current situation in a way that is understandable and interpretable by users.** Local explanations help even further by showing the reasons behind specific decisions. Local explanations support users to verify the accuracies of the decisions made by the automation [26]. Displays should facilitate a transparent view into the automated system's decision-making process by providing real-time feedback about why and how the automated system is currently making decisions [26], [66]. Real-time feedback in the display should illustrate how current inputs are being processed into outputs. This at least includes a clear depiction of the data being analyzed, the methods used to compute the outputs, and the system's confidence in the results produced. Such feedback can help users to monitor and understand the system's operations as they occur.

Understanding how and why a particular decision was made by an automated system for a specific instance or observation is important [67]. When automated outputs are similar to users' expectations, understanding the computations can help mitigate overconfidence and break fixation on the automated solution. When automated outputs are dissimilar from users' expectations, understanding the computations can help users reconcile differences in stance and perspective. Over time, continued experience with the automated system's computations can help users develop a better understanding of how the automated system operates in general so that they can better understand specific instances in the future.

*C. Clarity in System Modes and Mode Changes (C12)*

**Displays should convey the automated system's current mode, how it functions in this mode, the functionality and goals inherent to the mode, and the conditions under which the system will transition to another mode.** In an automated system, a mode is an operational state where the system has a given set of behaviors, goals, and priorities. For instance, a smartphone may come in 'airplane mode' or 'do not disturb mode', each one changing the functionality of the device and how it reacts to inputs. Understanding the current mode of automation consists of at least the goals and priorities of the automated system in this mode, how the automated system computes outputs from inputs in this mode, and under what conditions the automated system will switch to a different mode.

Mode change is the procedure through which the automated system transitions from one operating state to another. This might be triggered by user commands, pre-defined conditions being satisfied, or due to stimuli from the environment. Mode changes can be confusing or surprising if not clearly communicated [68]. A good mode change should give clear notice to the user, consistently indicate the current mode in a way that is permanently visible to the user, transition between modes without causing system instability, and take measures to avoid unintended behavior. Automated systems will often have multiple modes of operation. In these different modes, interacting with what seems to be the same display and same inputs can result in meaningfully different behaviors or outputs from the automated system [63]. Users need to understand the impacts of automated systems operating in different modes [11].

*D. Predictable Automation Behavior (C13)*

**Displays should enable users to reasonably predict the future actions of the automated systems and how they change over time.** Users cannot effectively coordinate activities with unpredictable automation. Displays should be designed not just to inform users about the current state and actions of automated systems but also to provide cues that allow users to predict the future actions of the automated system to a reasonable degree of accuracy [35], [67]. This includes predicting current inputs and outputs of the automated system, what future inputs are likely to be based on current inputs, how the automated system is likely to compute future outputs based on those expected future inputs, and potentially how the automated system itself is changing over time. This predictive capability is essential for effective human-machine coordination [20]. Otherwise, users are likely to be surprised by automation outputs and reactive (rather than proactive) in their response [63]. In high tempo work, users may not have time to recover from these automation surprises.

*E. Conveying System Anomaly and Misalignment (C14)*

**Displays must convey to users the ways an automated system is misaligned to the world, even when the automated system is unaware it is misaligned to the world.** Automation cannot reliably determine when it is misaligned or miscalibrated to the world; instead, other agents (e.g., people) must be able to recognize when and how the automation is misaligned to the world [46], [64], [67], so they can take appropriate actions to recalibrate or work around it.

At minimum, the display must convey the current state of the world, how and why the automated system is interpreting the current state of the world, and how this interpretation is misaligned to the world itself [64]. This is best achieved by setting the automation's outputs in the context of related information which may give clues as to when and how the automation may be misaligned, including: the data currently used by the automated system relative to the data available in the world [45], the automation's current goal prioritization relative to alternatives [69], and the automation's model of the world [56].

To understand this heuristic better, we make a distinction between automation anomalies and data anomalies. Automation anomalies occur when the internal models or processing of automation are not in synchronization with the environment. It can be due to a software bug, an outdated algorithm, mismatch in calibration, or the algorithm failing to fully capture the richness and complexity of the real-world. Data anomalies are those that involve unexpected patterns within the data being processed by the system that could indicate new trends, outliers, or errors in data. Misalignment happens when the automated system's operations and the real-world conditions do not match. For example, if an autonomous vehicle's sensors become obscured by mud, the vehicle's perception system might misinterpret a clear road with for an obstacle-loaded path. Another example could be a smart irrigation system that fails to adjust to changing weather patterns due to outdated algorithms or lack of data, resulting in over-watering during a rainy season.

## VIII. Discussion

In this work, we revised prior guidelines [18] to propose fourteen heuristics designed to augment support for five key macrocognitive functions: event detection, sensemaking, adaptability, perspective shifting, and coordination. These heuristics are critical in joint human-machine cognitive system designs aiming to maximally improve operational performance and interaction with users. They operationalize a structure for addressing the intricacies of human-machine interaction to make the user experience more intuitive and responsive.

This set of fourteen heuristics is intended to create a better balance between academic rigor and practical applicability to operational needs of joint human-machine cognitive systems. They were written to resonate with practical experiences and mental models of operational personnel that help in forming a deeper connection and understanding, thus making theoretical insights more actionable in real-world applications [19]. However, we were particularly mindful to avoid oversimplifying the inherent complexity of such interactions in the synthesis of the extensive literature on joint human-machine cognitive systems. Each heuristic was developed to evoke an appreciation of the complexity involved to support operational personnel in managing intricate cognitive tasks. This methodological choice aligns with our goal of making the research on joint cognitive systems broadly impactful [19], [70].

However, it is important for these heuristics to be empirically evaluated in practice. Such assessments are critical for determining the practical utility, validity, and overall effectiveness of these heuristics. It is only through administering a test such as this that we can assess how effectively these heuristics help to empower people in organizations to design and evaluate technologies. In attempting to close the gap between theoretical research and its practical application, it is important for future research to conduct practical assessments.

We believe these heuristics are a step towards empowering people in organizations to not only improve the design and evaluation of displays but also to foster a more intuitive and efficient interaction between humans and machines. Future research should continue to improve these heuristics based on data and feedback in the field so that joint cognitive system design processes can be optimized to have a large-scale impact [19]. In essence, this discussion sets the basis for future continued dialogue and development in cognitive systems engineering. Ultimately, the goal of these heuristics is to not only guide but also enhance the design and usability of complex systems, leading to more effective human-machine collaborations.

## Acknowledgment

This material is based upon work supported by the United States Air Force under Contract No. FA8650-22-C-6453 and FA8650-22-F-6403. Any opinions, findings, and conclusions or recommendations expressed in this material are those of the authors and do not necessarily reflect the views of the United States Air Force. We would like to thank those who have contributed to the development and advancement of these heuristics, including Aaron Cochran, Prerana Walli, Kenneth Cassidy, Priyanka Tewani, Morgan Reynolds, Samantha Malone, Nicolette McGeorge, and Taylor Murphy.

## References

[1] B. Shneiderman, "Bridging the Gap Between Ethics and Practice: Guidelines for Reliable, Safe, and Trustworthy Human-centered AI Systems," *ACM Trans Interact Intell Syst*, vol. 10, no. 4, p. 26:1-26:31, Oct. 2020, doi: 10.1145/3419764.

[2] M. F. Rayo, "Designing for collaborative autonomy: updating user-centered design heuristics and evaluation methods," *Proc. Hum. Factors Ergon. Soc. Annu. Meet.*, vol. 61, no. 1, pp. 1091–1095, Oct. 2017, doi: 10.1177/1541931213601877.

[3] J. Nielsen, "Enhancing the explanatory power of usability heuristics," *Proc. SIGCHI Conf. Hum. Factors Comput. Syst. Celebr. Interdepend.*, pp. 152–158, 1994.

[4] G. Klein, K. G. Ross, B. M. Moon, D. E. Klein, R. R. Hoffman, and E. Hollnagel, "Macrocognition," *IEEE Intell. Syst.*, vol. 18, no. 3, pp. 81–85, May 2003, doi: 10.1109/MIS.2003.1200735.

[5] R. Hoffman, "An Integrated Model of Macrocognitive Work and Trust in Automation," 2013.

[6] R. R. Hoffman and M. D. McNeese, "A History for Macrocognition," *J. Cogn. Eng. Decis. Mak.*, vol. 3, no. 2, pp. 97–110, Jun. 2009, doi: 10.1518/155534309X441835.

[7] K. L. Mosier *et al.*, "Automation, Task, and Context Features: Impacts on Pilots' Judgments of Human–Automation Interaction," *J. Cogn. Eng. Decis. Mak.*, vol. 7, no. 4, pp. 377–399, Dec. 2013, doi: 10.1177/1555343413487178.

[8] N. B. Sarter and D. D. Woods, "Pilot Interaction With Cockpit Automation II: An Experimental Study of Pilots' Model and Awareness of the Flight Management System," *Int. J. Aviat. Psychol.*, vol. 4, no. 1, pp. 1–28, Jan. 1994, doi: 10.1207/s15327108ijap0401_1.

[9] B. Moon, R. Hoffman, M. Lacroix, E. Fry, and A. Miller, "Exploring Macrocognitive Healthcare Work: Discovering Seeds for Design Guidelines for Clinical Decision Support," presented at the

International Conference on Applied Human Factors and Ergonomics (AHFE), 2021. doi: 10.54941/ahfe100532.
[10] S. Amershi *et al.*, "Guidelines for Human-AI Interaction," in *Proceedings of the 2019 CHI Conference on Human Factors in Computing Systems*, in CHI '19. Glasgow Scotland Uk: Association for Computing Machinery, May 2019, pp. 1–13. doi: 10.1145/3290605.3300233.
[11] M. R. Endsley, "From Here to Autonomy: Lessons Learned From Human–Automation Research," *Hum. Factors J. Hum. Factors Ergon. Soc.*, vol. 59, no. 1, pp. 5–27, Feb. 2017, doi: 10.1177/0018720816681350.
[12] K. M. Feigh and A. R. Pritchett, "Requirements for Effective Function Allocation: A Critical Review," *J. Cogn. Eng. Decis. Mak.*, vol. 8, no. 1, pp. 23–32, Mar. 2014, doi: 10.1177/1555343413490945.
[13] M. Johnson, J. M. Bradshaw, R. R. Hoffman, P. J. Feltovich, and D. D. Woods, "Seven Cardinal Virtues of Human-Machine Teamwork: Examples from the DARPA Robotic Challenge," *IEEE Intell. Syst.*, vol. 29, no. 6, pp. 74–80, Nov. 2014, doi: 10.1109/MIS.2014.100.
[14] P. J. Smith and R. R. Hoffman, Eds., *Cognitive Systems Engineering: A Future for a Changing World*. in Expertise: research and applications. Boca Raton London New York: CRC Press, Taylor & Francis Group, 2017.
[15] D. D. Woods, "Essentials of resilience, revisited," in *Handbook on Resilience of Socio-Technical Systems*, M. Ruth and S. Goessling-Reisemann, Eds., in *Handbook on Resilience of Socio-Technical Systems*. , United Kingdom: Edward Elgar Publishing, 2019, pp. 52–65. doi: 10.4337/9781786439376.00009.
[16] S. W. A. Dekker and D. D. Woods, "MABA-MABA or Abracadabra? Progress on Human–Automation Co-ordination," *Cogn. Technol. Work*, vol. 4, no. 4, pp. 240–244, Nov. 2002, doi: 10.1007/s101110200022.
[17] G. Baxter and I. Sommerville, "Socio-technical systems: From design methods to systems engineering," *Interact. Comput.*, vol. 23, no. 1, pp. 4–17, Jan. 2011, doi: 10.1016/j.intcom.2010.07.003.
[18] D. A. Morey *et al.*, "Towards Joint Activity Design Heuristics: Essentials for Human-Machine Teaming," *Proc. Hum. Factors Ergon. Soc. Annu. Meet.*, p. 21695067231193646, Oct. 2023, doi: 10.1177/21695067231193646.
[19] M. C. Fitzgerald, "The IMPActS Framework: the necessary requirements for making science-based organizational impact," The Ohio State University, 2019. Accessed: Jun. 01, 2022. [Online]. Available: https://etd.ohiolink.edu/apexprod/rws_olink/r/1501/10?clear=10&p10_accession_num=osu1557191348657812
[20] G. Klein, D. D. Woods, J. M. Bradshaw, R. R. Hoffman, and P. J. Feltovich, "Ten Challenges for Making Automation a 'Team Player' in Joint Human-Agent Activity," *IEEE Intell. Syst.*, vol. 19, no. 6, pp. 91–95, Nov. 2004, doi: 10.1109/MIS.2004.74.
[21] M. R. Endsley, "Toward a Theory of Situation Awareness in Dynamic Systems," *Hum. Factors J. Hum. Factors Ergon. Soc.*, vol. 37, no. 1, pp. 32–64, 1995, doi: 10.1518/001872095779049543.
[22] E. Hollnagel and D. D. Woods, *Joint Cognitive Systems: Foundations of Cognitive Systems Engineering*, 0 ed. CRC Press, 2005. doi: 10.1201/9781420038194.
[23] R. R. Hoffman, S. T. Mueller, G. Klein, and J. Litman, "Metrics for Explainable AI: Challenges and Prospects," 2018, *arXiv*: arXiv:1812.04608. Accessed: Oct. 26, 2022. [Online]. Available: http://arxiv.org/abs/1812.04608
[24] G. Klein, P. J. Feltovich, J. M. Bradshaw, and D. D. Woods, "Common Ground and Coordination in Joint Activity," in *Organizational Simulation*, W. B. Rouse and K. R. Boff, Eds., Hoboken, NJ, USA: John Wiley & Sons, Inc., 2005, pp. 139–184. doi: 10.1002/0471739448.ch6.
[25] M. L. Bolton, J. Edworthy, and A. D. Boyd, "A Formal Analysis of Masking Between Reserved Alarm Sounds of the IEC 60601-1-8 International Medical Alarm Standard," *Proc. Hum. Factors Ergon. Soc. Annu. Meet.*, vol. 62, no. 1, pp. 523–527, Sep. 2018, doi: 10.1177/1541931218621119.
[26] R. R. Hoffman, M. Jalaeian, C. Tate, G. Klein, and S. T. Mueller, "Evaluating machine-generated explanations: a 'Scorecard' method for XAI measurement science," *Front. Comput. Sci.*, vol. 5, 2023, Accessed: Nov. 30, 2023. [Online]. Available: https://www.frontiersin.org/articles/10.3389/fcomp.2023.1114806
[27] M. F. Rayo, E. S. Patterson, M. Abdel-Rasoul, and S. D. Moffatt-Bruce, "Using timbre to improve performance of larger auditory alarm sets," *Ergonomics*, vol. 62, no. 12, pp. 1617–1629, Dec. 2019, doi: 10.1080/00140139.2019.1676473.
[28] E. S. Patterson, E. M. Roth, and D. D. Woods, "Predicting Vulnerabilities in Computer-Supported Inferential Analysis under Data Overload," *Cogn. Technol. Work*, vol. 3, no. 4, pp. 224–237, Dec. 2001, doi: 10.1007/s10111-001-8004-y.
[29] D. D. Woods, E. S. Patterson, and E. M. Roth, "Can We Ever Escape from Data Overload? A Cognitive Systems Diagnosis," *Cogn. Technol. Work*, vol. 4, no. 1, pp. 22–36, Apr. 2002, doi: 10.1007/s101110200002.
[30] E. S. Patterson, D. D. Woods, D. Tinapple, and E. M. Roth, "Using Cognitive Task Analysis (CTA) to Seed Design Concepts for Intelligence Analysts Under Data Overload," *Proc. Hum. Factors Ergon. Soc. Annu. Meet.*, vol. 45, no. 4, pp. 439–443, Nov. 2001, doi: 10.1177/154193120104500437.
[31] J. J. Schlesinger *et al.*, "Acoustic features of auditory medical alarms—An experimental study of alarm volume," *J. Acoust. Soc. Am.*, vol. 143, no. 6, pp. 3688–3697, Jun. 2018, doi: 10.1121/1.5043396.
[32] K. B. Bennett, A. L. Nagy, and J. M. Flach, *Visual Displays*. in Handbook of Human Factors and Ergonomics. John Wiley & Sons, Inc., 2012. doi: 10.1002/9781118131350.ch42.
[33] K. B. Bennett and J. M. Flach, "Graphical Displays: Implications for Divided Attention, Focused Attention, and Problem Solving," *Hum. Factors J. Hum. Factors Ergon. Soc.*, vol. 34, no. 5, pp. 513–533, Oct. 1992, doi: 10.1177/001872089203400502.
[34] J. Holt, K. B. Bennett, and J. M. Flach, "Emergent features and perceptual objects: re-examining fundamental principles in analogical display design," *Ergonomics*, vol. 58, no. 12, pp. 1960–1973, Dec. 2015, doi: 10.1080/00140139.2015.1049217.
[35] J. M. Bradshaw, P. J. Feltovich, and M. Johnson, "Human-Agent Interaction," in *The handbook of human-machine interaction*, CRC Press, 2011, pp. 283–300. [Online]. Available: https://www.researchgate.net/publication/267819585_Human-Agent_Interaction
[36] D. D. Woods, "Toward a theoretical base for representation design in the computer medium: Ecological perception and aiding human cognition," in *An Ecological Approach To Human Machine Systems I: A Global Perspective*, J. M. Flach, P. A. Hancock, J. Caird, and K. J. Vicente, Eds., Erlbaum, 1995.
[37] G. A. Klein, J. K. Phillips, E. L. Rall, and D. A. Peluso, "A Data-Frame Theory of Sensemaking," in *Expertise Out of Context*, R. R. Hoffman, Ed., in Expertise Out of Context: Proceedings of the Sixth International Conference of Naturalistic Decision Making. , Psychology Press, 2007. [Online]. Available: http://books.google.com/books?hl=en\&lr=\&id=GQK8mccWlVoC\&oi=fnd\&pg=PA113\&ots=cW8D\_pN7TA\&sig=Q6lUlb\_-9YPc2rxJ1gLXCzSzsUI\#v=onepage\&q\&f=false
[38] R. R. Hoffman and G. Klein, "Explaining Explanation, Part 1: Theoretical Foundations," *IEEE Intell. Syst.*, vol. 32, no. 3, pp. 68–73, May 2017, doi: 10.1109/MIS.2017.54.
[39] C. M. Burns and J. R. Hajdukiewicz, *Ecological Interface Design*. in CRC. CRC, 2004.
[40] E. S. Patterson, E. M. Roth, and D. D. Woods, "Facets of Complexity in Situated Work," in *Macrocognition Metrics and Scenarios : Design and Evaluation for Real-World Teams*, E. S. Patterson and J. E. Miller, Eds., in Macrocognition Metrics and Scenarios. , Taylor & Francis Group, 2018, pp. 221–251. [Online]. Available: http://www.worldcat.org/title/macrocognition-metrics-and-scenarios-design-and-evaluation-for-real-world-teams/oclc/754793942
[41] M. F. Rayo, "Directive Displays: Supporting Human-machine Coordination by Dynamically Varying Representation, Information, and Interjection Strength," Dissertation, The Ohio State University, 2013. [Online]. Available: http://rave.ohiolink.edu/etdc/view?acc_num=osu1373384199
[42] D. D. Woods and E. Hollnagel, "Prologue: Resilience Engineering Concepts," in *Resilience Engineering: Concepts and Precepts*, 1st ed., E. Hollnagel, D. D. Woods, and N. Leveson, Eds., CRC Press, 2006, pp. 1–6. doi: 10.1201/9781315605685-1.
[43] J. R. Edworthy *et al.*, "Getting Better Hospital Alarm Sounds Into a Global Standard," *Ergon. Des. Q. Hum. Factors Appl.*, vol. 26, no. 4, pp. 4–13, Oct. 2018, doi: 10.1177/1064804618763268.

[44] M. Watson and P. Sanderson, "Sonification supports eyes-free respiratory monitoring and task time-sharing," *Hum. Factors J. Hum. Factors Ergon. Soc.*, vol. 46, no. 3, pp. 497–517, 2004.

[45] M. F. Rayo, C. R. Horwood, M. C. Fitzgerald, M. R. Grayson, M. Abdel-Rasoul, and S. D. Moffatt-Bruce, "Situated Visual Alarm Displays Support Machine Fitness Assessment for Nonexplainable Automation," *IEEE Trans. Hum.-Mach. Syst.*, vol. 52, no. 5, pp. 984–993, Oct. 2022, doi: 10.1109/THMS.2022.3155714.

[46] R. R. Hoffman, P. J. Feltovich, K. M. Ford, and D. D. Woods, "A rose by any other name... would probably be given an acronym [cognitive systems engineering]," *IEEE Intell. Syst.*, vol. 17, no. 4, pp. 72–80, Jul. 2002, doi: 10.1109/MIS.2002.1024755.

[47] E. S. Patterson, J. Watts-Perotti*, and D. D. Woods, "Voice Loops as Coordination Aids in Space Shuttle Mission Control," *Comput. Support. Coop. Work CSCW*, vol. 8, no. 4, pp. 353–371, Dec. 1999, doi: 10.1023/A:1008722214282.

[48] D. D. Woods and E. Hollnagel, *Joint Cognitive Systems: Patterns in Cognitive Systems Engineering*, 0 ed. CRC Press, 2006. doi: 10.1201/9781420005684.

[49] A. M. Madni and C. C. Madni, "Architectural Framework for Exploring Adaptive Human-Machine Teaming Options in Simulated Dynamic Environments," *Systems*, vol. 6, no. 4, p. 44, Dec. 2018, doi: 10.3390/systems6040044.

[50] M. F. Rayo *et al.*, "Interactive questioning in critical care during handovers: a transcript analysis of communication behaviours by physicians, nurses and nurse practitioners," *BMJ Qual. Saf.*, vol. 23, no. 6, pp. 483–489, Jun. 2014, doi: 10.1136/bmjqs-2013-002341.

[51] N. B. Sarter and B. Schroeder, "Supporting Decision Making and Action Selection under Time Pressure and Uncertainty: The Case of In-Flight Icing," *Hum. Factors*, vol. 43, no. 4, pp. 573–583, Dec. 2001, doi: 10.1518/001872001775870403.

[52] P. J. Smith, C. E. McCoy, and C. Layton, "Brittleness in the design of cooperative problem-solving systems: the effects on user performance," *IEEE Trans. Syst. Man Cybern. - Part Syst. Hum.*, vol. 27, no. 3, pp. 360–371, May 1997, doi: 10.1109/3468.568744.

[53] W. Elm, S. Potter, J. Tittle, D. Woods, J. Grossman, and E. Patterson, "Finding Decision Support Requirements for Effective Intelligence Analysis Tools," *Proc. Hum. Factors Ergon. Soc. Annu. Meet.*, vol. 49, no. 3, pp. 297–301, Sep. 2005, doi: 10.1177/154193120504900318.

[54] F. Flemisch, M. Heesen, T. Hesse, J. Kelsch, A. Schieben, and J. Beller, "Towards a dynamic balance between humans and automation: authority, ability, responsibility and control in shared and cooperative control situations," *Cogn. Technol. Work*, vol. 14, no. 1, pp. 3–18, 2012, doi: 10.1007/s10111-011-0191-6.

[55] S. W. A. Dekker and D. D. Woods, "Wrong, Strong, and Silent: What Happens when Automated Systems With High Autonomy and High Authority Misbehave?," *J. Cogn. Eng. Decis. Mak.*, vol. 18, no. 4, pp. 339–345, Dec. 2024, doi: 10.1177/15553434241240849.

[56] D. A. Morey and M. F. Rayo, "Situated Interpretation and Data: Explainability to Convey Machine Misalignment," *IEEE Trans. Hum.-Mach. Syst.*, vol. 54, no. 1, pp. 100–109, Feb. 2024, doi: 10.1109/THMS.2023.3334988.

[57] P. J. Smith, "Making Brittle Technologies Useful," in *Cognitive Systems Engineering*, 2017, p. 28.

[58] M. E. Reynolds, "Modulating Monitoring and Adapting Authority Relationships: A Multi-Industry Study of Fluent Synchronization to Support Resilient Coordination," The Ohio State University, 2024. Accessed: Dec. 12, 2024. [Online]. Available: https://etd.ohiolink.edu/acprod/odb_etd/r/etd/search/10?p10_accession_num=osu172400834465066&clear=10&session=103536767175895

[59] S. A. Guerlain *et al.*, "Interactive Critiquing as a Form of Decision Support: An Empirical Evaluation," *Hum. Factors J. Hum. Factors Ergon. Soc.*, vol. 41, no. 1, pp. 72–89, Mar. 1999, doi: 10.1518/001872099779577363.

[60] D. D. Woods, "Visual momentum: a concept to improve the cognitive coupling of person and computer," *Int. J. Man-Mach. Stud.*, vol. 21, no. 3, pp. 229–244, Sep. 1984, doi: 10.1016/S0020-7373(84)80043-7.

[61] D. D. Woods, "The theory of graceful extensibility: basic rules that govern adaptive systems," *Environ. Syst. Decis.*, vol. 38, no. 4, pp. 433–457, Dec. 2018, doi: 10.1007/s10669-018-9708-3.

[62] R. R. Hoffman and D. D. Woods, "Beyond Simon's Slice: Five Fundamental Trade-Offs that Bound the Performance of Macrocognitive Work Systems," *IEEE Intell. Syst.*, vol. 26, no. 6, pp. 67–71, Nov. 2011, doi: 10.1109/MIS.2011.97.

[63] N. B. Sarter, D. D. Woods, and C. E. Billings, "Automation Surprises," *Handb. Hum. Factors Ergon.*, vol. 2, 1997, doi: 10.1201/9780849375477.ch587.

[64] M. F. Rayo *et al.*, "The Need for Machine Fitness Assessment: Enabling Joint Human-Machine Performance in Consumer Health Technologies," *Proc. Int. Symp. Hum. Factors Ergon. Healthc.*, vol. 9, no. 1, pp. 40–42, 2020, doi: 10.1177/2327857920091041.

[65] R. R. Hoffman, S. T. Mueller, and G. Klein, "Explaining Explanation, Part 2: Empirical Foundations," *IEEE Intell. Syst.*, vol. 32, no. 4, pp. 78–86, 2017, doi: 10.1109/MIS.2017.3121544.

[66] J. Y. C. Chen and M. J. Barnes, "Agent Transparency for Human-Autonomy Teaming," in *Human-Automation Interaction*, vol. 12, V. G. Duffy, M. Ziefle, P.-L. P. Rau, and M. M. Tseng, Eds., in Automation, Collaboration, & E-Services, vol. 12. , Cham: Springer International Publishing, 2023, pp. 255–266. doi: 10.1007/978-3-031-10788-7_15.

[67] P. McDermott, C. Dominguez, N. Kasdaglis, M. Ryan, I. Trhan, and A. Nelson, "Human-Machine Teaming Systems Engineering Guide," MITRE CORP BEDFORD MA BEDFORD United States, Technical Report, Dec. 2018.

[68] N. B. Sarter and D. D. Woods, "How in the World Did We Ever Get into That Mode? Mode Error and Awareness in Supervisory Control," *Hum. Factors J. Hum. Factors Ergon. Soc.*, vol. 37, no. 1, pp. 5–19, Mar. 1995, doi: 10.1518/001872095779049516.

[69] M. F. Rayo, N. Kowalczyk, B. W. Liston, E. B.-N. Sanders, S. White, and E. S. Patterson, "Comparing the Effectiveness of Alerts and Dynamically Annotated Visualizations (DAVs) in Improving Clinical Decision Making," *Hum. Factors J. Hum. Factors Ergon. Soc.*, vol. 57, no. 6, pp. 1002–1014, Sep. 2015, doi: 10.1177/0018720815585666.

[70] C. Dominguez *et al.*, "Panel Discussion Cognitive Engineering: Will They Know Our Name When We Are 40?," *Proc. Hum. Factors Ergon. Soc. Annu. Meet.*, vol. 65, no. 1, pp. 257–261, Sep. 2021, doi: 10.1177/1071181321651042.